\documentclass[3p,times,twocolumn]{elsarticle}

\usepackage{ecrc}

\usepackage{graphicx}
\graphicspath{ {images/} }

\usepackage{refcount}

\volume{00}

\firstpage{1}

\journalname{Nuclear and Particle Physics Proceedings}

\runauth{}

\jid{nppp}

\jnltitlelogo{Nuclear and Particle Physics Proceedings}

\usepackage{amssymb}

\biboptions{sort&compress}

\usepackage[figuresright]{rotating}
\usepackage{amsmath}

\begin{document}

\begin{frontmatter}



\dochead{}

\title{$\Upsilon$ production in p-Pb and Pb-Pb collisions with ALICE at the LHC}


\author{Gabriele Gaetano Fronz\'e}
\address{For the ALICE Collaboration}
\address{INFN and University of Turin (IT), Subatech de Nantes (FR)}

\begin{abstract}
ALICE (A Large Ion Collider Experiment) is devoted to the study of heavy-ion collisions at LHC energies. In such collisions a deconfined state of nuclear matter, the Quark-Gluon Plasma (QGP), is formed. Due to their early production, quarkonium states are good probes to study the QGP evolution. Such states are affected by suppression mechanisms which lead to reduced yields with respect to pp and p-Pb collisions, while regeneration phenomena might lead to an enhancement of their production. The latter effects are expected to be negligible at LHC for bottomonium states. 
The recent ALICE results on $\Upsilon$ production in Pb-Pb collisions at $\sqrt{s_{\rm NN}}=5.02\;\rm{T}e\rm{V}$ will be presented and compared with previous measurements at $\sqrt{s_{\rm NN}}=2.76\;\rm{T}e\rm{V}$. A comparison with theoretical calculations will be performed as well. Results obtained in p-Pb collisions at $\sqrt{s_{\rm NN}}=5.02\;\rm{T}e\rm{V}$ will also be discussed.
\end{abstract}


\end{frontmatter}


\section{Motivations for quarkonium study}
\label{}
Quarkonium states are formed early during Quark Gluon Plasma (QGP \cite{Shuryak:1978ij}), hence they cross the whole QGP evolution probing the medium created in the collisions. Colour screening effects, sequential suppression and regeneration phenomena are the mechanisms affecting quarkonium production in the QGP \cite{Matsui:1986dk,Brambilla:2010cs,Cacciari:2012ny}.
Bottomonium mesons are bound states of b quark and antiquark. Bottomonium is a good candidate for the study of QGP since, with respect to lower mass quark bound states:
\begin{itemize}
        \item the perturbative theoretical approach is more reliable since bottom quark mass is higher;
	\item it presents no feed down from open bottom flavoured states;
	\item the regeneration is less relevant  \cite{Emerick:2011xu};
	\item Cold Nuclear Matter (CNM) effects are expected to be smaller.
\end{itemize} 
Moreover bottomonium states study is complementary to the study of charmonium states since they allow to study a different Bjorken-x range.
The modifications of quarkonium production yields, in heavy ion collisions, is evaluated through the nuclear modification factor $R_{\rm{AA}}$. The $R_{\rm{AA}}$, defined in \eqref{eq:RAA}, is the ratio between the production cross section measured in A-A collisions and in pp collisions ($\sigma_{\rm{pp}}$) rescaled by the nuclear overlap function $\langle T_{AA} \rangle$.
\begin{equation}
\label{eq:RAA}
\begin{gathered}
R_{AA}=\frac{N_{\rm{AA}}}{\langle T_{AA} \rangle\cdot\sigma_{\rm{pp}}}
\end{gathered}
\end{equation}
ALICE \cite{Aamodt:2008zz} has already published results in Pb-Pb collisions at $\sqrt{s_{\rm NN}}=2.76\;\rm{T}e\rm{V}$ \cite{Abelev:2014nua}. The $R_{\rm{AA}}$ has been computed using the $\sigma_{\rm{pp}}$ value evaluated using measurements by the LHCb Collaboration \cite{Aaij:2014nwa}. $\Upsilon(1\rm{S})$ results presented a centrality dependence of $R_{\rm{AA}}$, with stronger suppression in the most central events. ALICE results are also compatible with CMS $R_{\rm{AA}}$ measured at mid rapidity \cite{Chatrchyan:2012lxa}.

%

\section{Experimental setup}
\label{}
ALICE is composed by two groups of detectors: the central barrel and the muon spectrometer. Detectors from both groups have been used for this analysis. A detailed description of the whole apparatus can be found in \cite{Aamodt:2008zz,Abelev:2014ffa}.
The interaction vertex identification has ben performed using the two silicon pixel layers of the six layers of silicon detectors composing the Inner Tracking System (ITS) \cite{Aamodt:2010aa}.
The minimum bias trigger is provided by the V0 \cite{Abbas:2013taa}, a group of two arrays of scintillators, placed in the pseudo-rapidity ranges $2.8<\eta<5.1$ (V0-A) and $-3.7<\eta<-1.7$ (V0-C). The centrality estimation is obtained trough a Glauber fit of the V0 raw signal amplitudes \cite{Abbas:2013taa,Abelev:2013qoq}.
The two Zero Degree Calorimeters (ZDC) \cite{Abelev:2014ffa}, installed at $\pm114\;m$ from the Interaction Point in the accelerator tunnel, are used to reject electromagnetic events and to remove beam-induced background.
The muon spectrometer system \cite{Aamodt:2011gj} is located at $-4<\eta<-2.5$ and is specifically designed to track and identify muons. The innermost component is a ten radiation length thick front absorber. 
The muon tracker consists of five tracking stations, composed by two planes of cathode pad chambers each. It extends through a dipole which provides $3 \rm{T}\cdot\rm{m}$ integrated magnetic field to bend the charged particles trajectory. Downstream to the tracking system a 1.2 m thick (7.2 interaction lengths) iron wall stops efficiently the light hadrons coming from $\pi$ and K mesons decays. The muon trigger system \cite{Bossu:2012jt}, made of four planes of x-y reading RPC chambers, allows for online triggering on single muons and dimuons and for offline muon identification by matching with the tracker tracks \cite{Arnaldi:2006hi}. 
An additional absorber is placed around the beam line along the whole muon spectrometer length.

\section{Analysis strategy and data sample}
\label{}
The $\Upsilon(1\rm{S})$ yields are obtained fitting a $\mu^+\mu^{\--}$ invariant mass spectrum. The tracks of the muon tracker which are matched in the muon trigger are flagged and identified as muons. The muons used for the computation of the invariant mass spectrum are selected by applying cuts tuned to maximise the signal to background ratio. The cuts are:
\begin{itemize}
	\item $-4<\eta_\mu<-2.5$ to select muons within acceptance of the spectrometer;
	\item $p_{T \mu}\geq2\;\rm{GeV/c}$ to reduce combinatorial background;
	\item $17.6\;\rm{cm} < R_{\rm{abs}} < 89.5\;\rm{cm}$, where $R_{\rm{abs}}$ is the radial position of the track at the front absorber end, to reduce the contribution of particles from beam gas interactions.
\end{itemize}
A $-4<y_{\mu\mu}<-2.5$ cut is applied on the dimuon rapidity.
The fit of the invariant mass spectrum is performed with a function composed by the sum of one Extended Crystal Ball (CB2) for each resonance and a phenomenological background shape chosen among double or single exponentials or power laws.
The presented results have been obtained in p-Pb and Pb-Pb collisions at $\sqrt{s_{\rm NN}}=5.02\;\rm{T}e\rm{V}$. Since p-Pb collisions have been performed both with the proton or the Pb nucleus going towards the muon spectrometer, two rapidity ranges have been studied. The integrated luminosity values are reported in Table \ref{table:luminosity}.
The $\sigma_{pp}$ reference has to be measured at 
$\sqrt{s}=\sqrt{s_{\rm NN}}$ in order to compute the $R_{\rm{AA}}$ value. The low luminosity collected by ALICE in pp collisions at $\sqrt{s}=5.02\;\rm{T}e\rm{V}$ prevents the evaluation of the $\sigma_{\rm{pp}}$ reference, hence the cross section value has been computed by interpolating ALICE data at $\sqrt{s}=7$ and $8\;\rm{T}e\rm{V}$ \cite{Abelev:2014qha,Adam:2015rta} and LHCb data at $\sqrt{s}=2.76$, $7$ and $8\;\rm{T}e\rm{V}$ \cite{Aaij:2014nwa,Aaij:2015awa}. The interpolation method is described in detail in \cite{LHCb-CONF-2013-013}.
\begin{table}
\label{table:luminosity}
\begin{center}
  \begin{tabular}{| l | c | c | r }
    \hline
        Beam configuration & $\sqrt{s_{\rm NN}}$ & $L_{\rm{Int}}$ \\ \hline \hline
        p-Pb &$5.02\;\rm{T}e\rm{V}$& $5.0\rm{nb}^{-1}$ \\ 
        Pb-p &$5.02\;\rm{T}e\rm{V}$& $5.8\rm{nb}^{-1}$ \\ 
        Pb-Pb &$5.02\;\rm{T}e\rm{V}$& $225\rm{\mu b}^{-1}$ \\
    \hline
  \end{tabular}
\end{center}
\caption{Integrated luminosity for different beam configurations}
\end{table}
\section{p-Pb results}
\begin{figure}
    \centering
    \includegraphics[width=7cm]{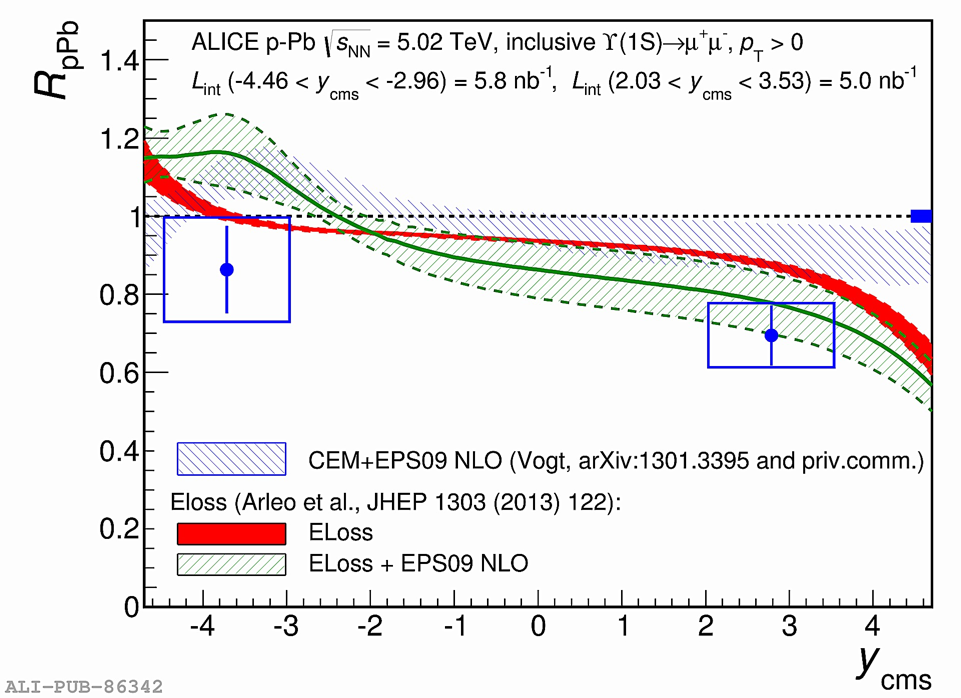}
    \caption{$\Upsilon$ $R_{\rm{pPb}}$ measured at $\sqrt{s_{\rm{NN}}}=5.02\;\rm{T}e\rm{V}$ compared with models as function of rapidity. \protect\footnotemark[\getrefnumber{note1}]}
    \label{fig:pPb_ALICE_rap_models}
\end{figure}

The p-Pb collisions are a CNM reference. Both forward ($2.03<y_{\rm{cms}}<3.53$) and backward ($-4.46<y_{\rm{cms}}<-2.96$) rapidity regions have been studied using inverse beam configurations.
At backward rapidity (Pb-going side) the $R_{\rm{pA}}$ values are compatible with no suppression, while at forward rapidity the results present a better agreement with models which foresee a reduction of the $\Upsilon$ yields \cite{Albacete:2013ei,Arleo:2012rs,Fujii:2013gxa,Ferreiro:2011xy}. The comparison of data with models suggests a better compatibility of experimental results with energy-loss only models at backward rapidity, while at forward rapidity the best agreement has been found with models containing both energy loss and NLO nuclear shadowing as shown in Fig.\ref{fig:pPb_ALICE_rap_models}. At backward rapidity the data suggests the models are overestimating the anti-shadowing contribution.

\section{Pb-Pb results}
\label{}
\begin{figure}
    \centering
    \includegraphics[width=7cm]{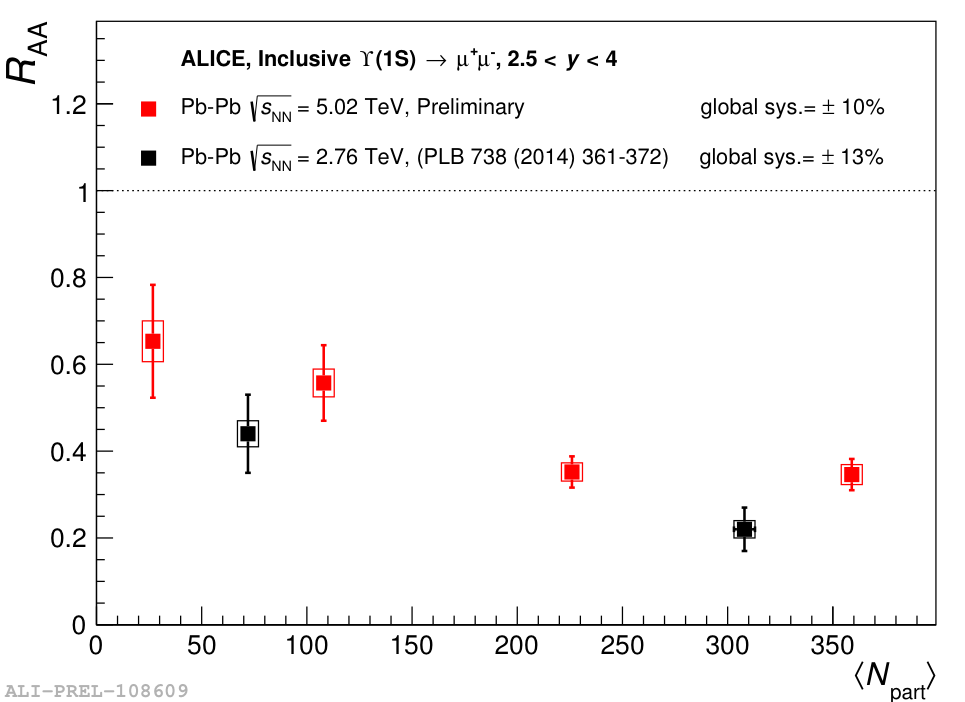}
    \caption{$\Upsilon$ $R_{\rm{AA}}$ measured in Pb-Pb at $\sqrt{s_{\rm{NN}}}=2.76\;\rm{T}e\rm{V}$ (black) and $\sqrt{s_{\rm{NN}}}=5.02\;\rm{T}e\rm{V}$ (red) represented as function of centrality. \protect\footnotemark[\getrefnumber{note1}]}
    \label{fig:PbPb_ALICE_cent}
\end{figure}

\begin{figure}
    \centering
    \includegraphics[width=7cm]{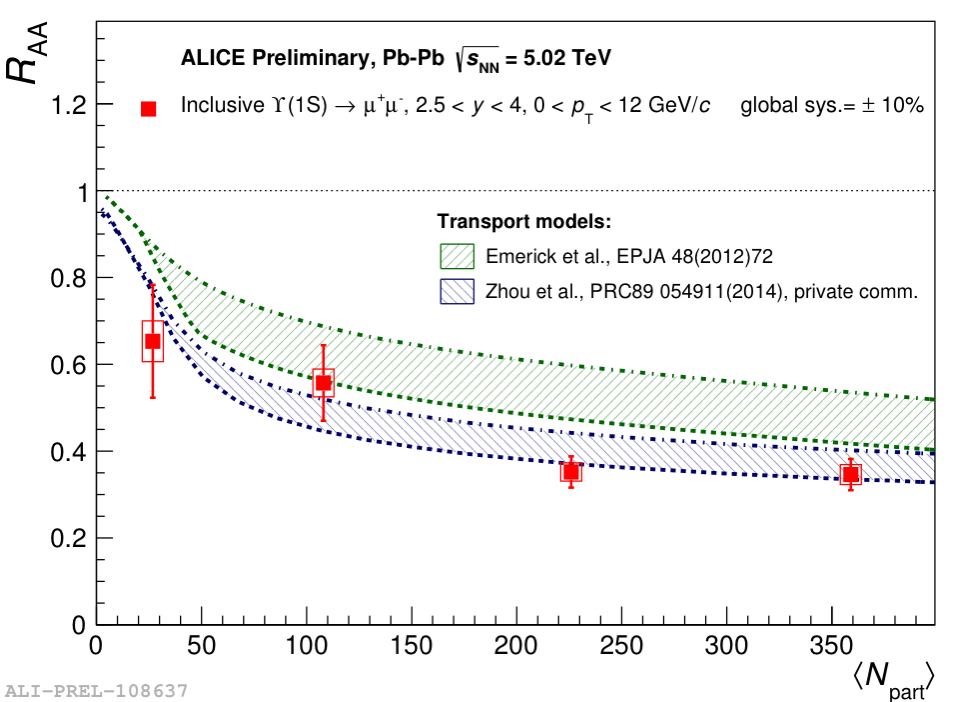}
    \caption{Measurements in Pb-Pb at  $\sqrt{s_{\rm{NN}}}=5.02\;\rm{T}e\rm{V}$ of the $\Upsilon$ $R_{\rm{AA}}$ represented as function of centrality and compared with models. \protect\footnotemark[\getrefnumber{note1}]}
    \label{fig:PbPb_ALICE_cent_models}
\end{figure}

\begin{figure}
    \centering
    \includegraphics[width=7cm]{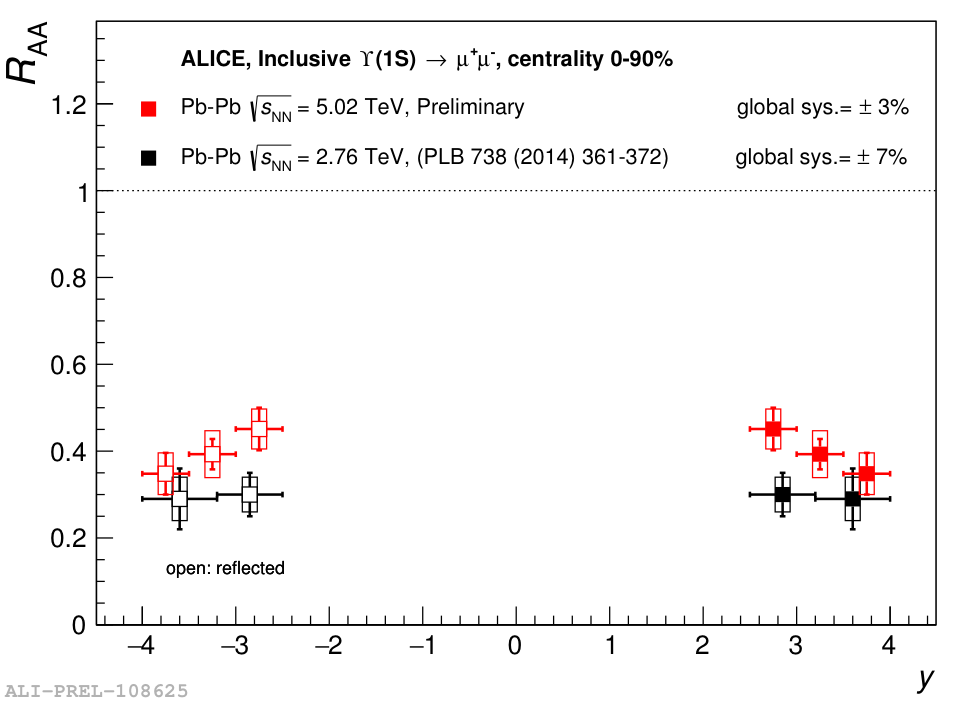}
    \caption{Measurements in Pb-Pb collisions at $\sqrt{s_{\rm{NN}}}=2.76 \;\rm{T}e\rm{V}$ (black) and $\sqrt{s_{\rm{NN}}}=5.02\;\rm{T}e\rm{V}$ (red) of the $\Upsilon$ $R_{\rm{AA}}$ represented as function of rapidity. \protect\footnotemark[\getrefnumber{note1}]}
    \label{fig:PbPb_ALICE_rap}
\end{figure}

\begin{figure}[h]
    \centering
    \includegraphics[width=7cm]{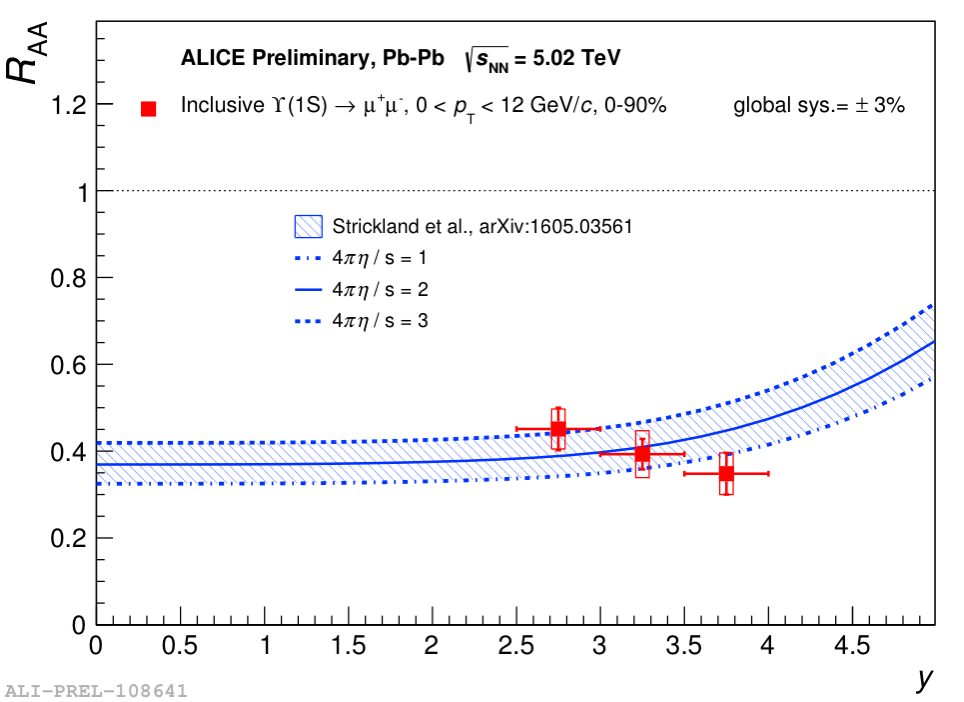}
    \caption{$R_{\rm{PbPb}}$ measured at  $\sqrt{s_{\rm{NN}}}=5.02\;\rm{T}e\rm{V}$ represented as function of rapidity and compared with models. \protect\footnotemark[\getrefnumber{note1}]}
    \label{fig:PbPb_ALICE_rap_models}
\end{figure}

The presented $R_{\rm{AA}}$ measurements at $\sqrt{s_{\rm{NN}}}=5.02\;\rm{T}e\rm{V}$ have been obtained analysing invariant mass spectra with a total of $N_{\Upsilon(1\rm{S})}=1107\pm70(stat.)\pm43(syst.)$ reconstructed $\Upsilon(1\rm{S})$ mesons, which corresponds to 10 times the $\Upsilon(1\rm{S})$ statistics collected at $\sqrt{s_{\rm{NN}}}=2.76\;\rm{T}e\rm{V}$. The systematic uncertainties are mainly due to signal extraction ($8-20\%$), $T_{\rm{AA}}$ evaluation ($1-3\%$) and tracker and trigger efficiencies ($4-7\%$).
The centrality dependence of $R_{\rm{AA}}$ is qualitatively similar to the one observed at $\sqrt{s_{\rm{NN}}}=2.76\;\rm{T}e\rm{V}$ as observed in Fig.\ref{fig:PbPb_ALICE_cent}. Even if the $R_{\rm{AA}}$ computed at $\sqrt{s_{\rm{NN}}}=5.02\;\rm{T}e\rm{V}$ is systematically above $R_{\rm{AA}}$ computed at $\sqrt{s_{\rm{NN}}}=2.76\rm{T}e\rm{V}$ the two values are compatible within uncertainties (see Fig.\ref{fig:PbPb_ALICE_cent}). 
The experimental data have been compared with two transport models (see Fig.\ref{fig:PbPb_ALICE_cent_models}). The Emerick model \cite{Emerick:2011xu} includes regeneration mechanisms, tuned on LHCb $b\bar b$ cross section measurement, and a feed-down contribution tuned on CDF data. The uncertainty bands have been obtained by varying the  nuclear shadowing amount from $0\%$ to $25\%$. The Zhou model \cite{PhysRevC.89.054911} includes no regeneration, but contains CNM effects tuned on EKS98 nuclear PDFs. The uncertainty bands are obtained by varying the feed-down fractions. Both the models compared to data are qualitatively capable of reproducing the observed trend within their uncertainty bands. With the current results no firm conclusion can be given about the presence of regeneration mechanism.
The rapidity dependence of $R_{\rm{AA}}$ has been measured. The trend of $R_{\rm{AA}}$ is growing from higher to lower $y$ (Fig.\ref{fig:PbPb_ALICE_rap}). The $R_{\rm{AA}}$ values at the two studied energies ($\sqrt{s_{\rm{NN}}}=2.76$ and $5.02\;\rm{T}e\rm{V}$) are compatible within uncertainties. The $R_{\rm{AA}}$ values are compared with Strickland model \cite{Krouppa:2016jcl} (See Fig.\ref{fig:PbPb_ALICE_rap_models}). The model foresees no regeneration or CNM  and includes hydrodynamic effects such as thermal suppression and anisotropic screening. The uncertainty bands are obtained through the variation of $\eta/s$\footnote{shear viscosity-to-entropy density ratio} ratio. Even if the slope suggested by experimental data seems to be opposite with respect to the one the model suggests, the agreement is still satisfied within uncertainties.

\setcounter{footnote}{2}
\footnotetext{\label{note1}The bars represent the statistical uncertainties, the boxes around the points the systematic ones, while the box drawn at $R_{\rm{pA}}$ or $R_{\rm{AA}}$ represents the global uncertainty.}

\section{Conclusions}
\label{}
The p-Pb analysis provided no significant observation of suppression at backward rapidity, while at forward rapidity a hint of suppression of the $\Upsilon(1\rm{S})$ production has been observed. All the tested models can reproduce within uncertainties the experimental data.
In the Pb-Pb analysis a strong centrality dependence of the $R_{\rm{AA}}$ has been observed, with smaller $R_{\rm{AA}}$ at higher centralities. No firm conclusion can be given about the energy hierarchy since the data points at $\sqrt{s_{\rm{NN}}}=2.76$ and $5.02\;\rm{T}e\rm{V}$ are compatible within uncertainties.
Some tension on the rapidity $R_{\rm{AA}}$ dependence between data and models has been observed, nevertheless the size of experimental and theoretical uncertainties prevents firm conclusions.




\nocite{*}
\bibliographystyle{elsarticle-num}
\bibliography{Fronze_G.bib}







\end{document}